\documentclass[aps,prb,twocolumn,superscriptaddress,floatfix,letter]{revtex4}
\usepackage{graphicx}
\usepackage{amssymb}   
\usepackage{amsfonts}
\usepackage{amsmath}
\usepackage{dsfont}
\usepackage{dcolumn}   
\usepackage{ulem}
\usepackage{bm}
\usepackage[mathscr]{eucal}
\usepackage[dvipsnames]{xcolor}
\usepackage[colorlinks, linkcolor=OrangeRed,citecolor=RoyalBlue,urlcolor=NavyBlue]{hyperref}
\usepackage[all]{hypcap} 
\usepackage{mathtools}
\usepackage{wasysym}

\definecolor{mypurple}{rgb}{0.49,0.18,0.56}
\definecolor{mygold}{rgb}{0.93,0.69,0.13}

\DeclareMathOperator{\Tr}{Tr}

\begin{document}

\title{Unconventional critical exponents at dynamical quantum phase transitions in a random Ising chain}
\author{Daniele Trapin}
\affiliation{Max-Planck-Institut f\"ur Physik komplexer Systeme, N\"othnitzer Stra{\ss}e 38,  01187-Dresden, Germany}
\author{Jad C. Halimeh}
\affiliation{INO-CNR BEC Center and Department of Physics, University of Trento, Via Sommarive 14, I-38123 Trento, Italy}
\author{Markus Heyl}
\affiliation{Max-Planck-Institut f\"ur Physik komplexer Systeme, N\"othnitzer Stra{\ss}e 38,  01187-Dresden, Germany}

\begin{abstract}
Dynamical quantum phase transitions (DQPTs) feature singular temporal behavior in transient quantum states during nonequilibrium real-time evolution.
In this work we show that DQPTs in random Ising chains exhibit critical behavior with nontrivial exponents that are not integer valued and not of mean-field type. 
By means of an exact renormalization group transformation we estimate the exponents with high accuracy eliminating largely any finite-size effects.
We further discuss how the considered dynamical phenomena can be made accessible in current Rydberg atom platforms.
In this context we explore signatures of the DQPTs in the statistics of spin configuration measurements available in such architectures.
Specifically, we study the statistics of clusters of consecutively aligned spins and observe a marked influence of the DQPT on the corresponding distribution.
\end{abstract}

\maketitle
\section{Introduction} \label{Introduction}
The advances in quantum simulators over the last decade have provided experimental access to the real-time dynamics of quantum matter at an unprecedented level of control.
This has led to the observation of many-body localization,\cite{schreiber2015,smith2016,Choi2016} time crystals,\cite{choi2017observation,zhang2017observation,Rovny2018,Smits2018} the quantum Kibble-Zurek mechanism,\cite{2014Xu,2016Anquez,2016Clark,Cui2016,keesling2019quantum,2020Xue} dynamics in gauge theories,\cite{martinez2016,Goerg2019,Schweizer2019,Mil2019,Yang2020} prethermalization,\cite{Gring2012,Langen2015,Neyenhuis2017,2019Singh} and various concepts of dynamical phase transitions.\cite{jurcevic2016,Flaeschner2017,Zhang2017,Smale2019,Yang2019,Tian2020}
A fundamental property of the quantum states generated through a nonequilibrium process is that they cannot be captured with a conventional thermodynamic description.
As a consequence, elementary equilibrium concepts such as phases and phase transitions require a generalization to the dynamical realm.

In this context, Dynamical Quantum Phase Transitions (DQPTs) have been introduced as an attempt to lift the notion of phase transitions and criticality to the dynamical regime.\cite{heyl2013,Heyl2017Review,Zvyagin2017Review,heyl2019dynamical,Mori_review}
As opposed to conventional transitions, which are driven by external control parameters, these DQPTs are signaled by singular behavior as a function of time and are therefore occurring due to drastic internal changes as a system evolves temporally.
In some cases it has been rigorously shown that such DQPTs can follow the equilibrium paradigm of continuous phase transitions.\cite{ScalingHeyl,karrasch2013,karrasch2017,homrighausen2017,lang2018,lang2018dynamical,PhysRevB.97.174303,zunkovic2016,Halimeh2018a,Hashizume2018,Hashizume2019}
However, the found associated critical exponents have typically been integer valued or of mean-field type.\cite{wu2019dynamical,wu2020dynamical,wu2020nonequilibrium}
Thus, it has remained as a central open question whether some generic quantum models exist exhibiting DQPTs with critical behavior featuring nontrivial exponents belonging to more exotic universality classes.

It is the central goal of this work to show that DQPTs in random one-dimensional Ising chains with bond disorder show critical behavior associated with a nontrivial exponent.
Using an exact real-space renormalization group treatment, we find that the dynamical analog $\lambda(t)$ of a free energy density follows a temporal scaling form $\lambda(t) \sim |(t - t_c)/t_c|^\alpha$ close to a DQPT at time $t_c$ with $\alpha=0.1264(2)$.
We argue that the considered real-time scenario is accessible with current experiments in Rydberg atoms trapped in optical tweezers,\cite{bernien2017probing,browaeys2020many,labuhn2016tunable, kim2018detailed, barredo2015coherent} where the random couplings in the Ising chain can be created by a suitable random spatial arrangement of the atoms in real-space.
In this context we further explore whether signatures of the DQPTs in experimentally accessible quantities other than $\lambda(t)$ can be observed.
A particular feature of these experiments is that single shots of a measurement yield spin configurations of all the individual Rydberg atoms, whose statistics we study via the occurrence of clusters with $M$ consecutive aligned spins.
We find that the associated probability distribution function $p(M,t)$ exhibits a distinct temporal signature of DQPTs.

%
It is a key challenge in the field of nonequilibrium real-time dynamics of quantum matter to characterize transient quantum states as they cannot be described in terms of ensembles as it is the case in equilibrium or for many steady states appearing in the asymptotic long-time limit.
In this context, the theory of DQPTs has been introduced~\cite{heyl2013,heyl2014dynamical} as a concept to provide a general framework for the identification of dynamical phases and their mutual transitions even without the possibility of an ensemble description.
The central quantity within this theory is the Loschmidt amplitude
\begin{equation}
    \mathcal{L}(t) = \langle \psi_0| e^{-iHt}| \psi_0 \rangle,
    \label{loschmidt}
\end{equation}
which is nothing but the overlap between the initial state $|\psi_0\rangle$ before and its time-evolved version $|\psi(t) \rangle = \exp(-iHt) |\psi_0\rangle$ after the quench.
On a formal level, $\mathcal{L}(t)$ assumes the form of a complex partition function. As a natural consequence, it is natural to introduce an effective free energy density (also termed rate function):
\begin{equation}
    \lambda(t)=-\frac{1}{N} \log(|\mathcal{L}(t)|^2).
    \label{lambda}
\end{equation}
In close analogy to equilibrium, where phase transitions are associated with singular behavior in free energies, a DQPT occurs whenever $\lambda(t)$ becomes nonanalytic.
Here, however, this nonanalytic behavior happens as a function of time and therefore as a consequence of drastic internal changes during the dynamics and not as a function of an external control parameter, as it is the case for equilibrium transitions.
In the meantime it has been explored extensively to which extent properties of conventional phase transitions can be generalized to DQPTs.\cite{Heyl2017Review}
It is of particular importance, that DQPTs can follow the equilibrium paradigm of continuous phase transitions involving scaling and universality.
Concretely, this has been shown for one-dimensional Ising chains using exact renormalization group transformations where the exact fixed points and their critical behavior have been identified.\cite{ScalingHeyl}
While for many models a close analogy between the equilibrium phase diagrams and DQPTs have been observed,\cite{budich2016dynamical,Dutta2017} also exceptions have been found such as in the case of long-range Ising models,\cite{halimeh2017dynamical,zauner2017probing} where the corresponding DQPTs have been termed anomalous.
%
Importantly, DQPTs have not only remained a theoretical framework, but it has also become of significant experimental interest. DQPTs have been observed in a trapped ion experiment,\cite{jurcevic2016} where the dynamics of the transverse Ising model was simulated and it was possible to measure the rate function $\lambda(t)$.\cite{jurcevic2016}
Further, DQPTs have been explored in systems of ultra-cold atoms,\cite{Flaeschner2017} quantum walks,\cite{xu2020measuring,wang2019simulating} nitrogen-vacancy centers in diamond,\cite{wang2019experimental,Yang2019} topological nanomechanical systems,\cite{Tian2019} and superconducting qubits.\cite{Guo2019}
This manuscript is organized as follows. In Sec.~\ref{model} we introduce the main model we use for our analysis. In Sec.~\ref{critical_exponent} we compute the Loschmidt echo in terms of the complex Ising partition function, and then numerically extract the critical exponent associated with the nonanalytic cusps arising in the Loschmidt return rate.
In Sec.~\ref{cluster_size} we present an experimentally inspired scheme in order to detect signatures of DQPTs based on projective measurements of spin configurations. Also in the spirit of experimental relevance, we add in Sec.~\ref{perturbative} a weak random longitudinal field in our model to investigate the effect of possible noise on our results. We finally conclude and propose future investigations in Sec.~\ref{Conclusions}.

\section{Model and quench} \label{model}
In this work we consider the one-dimensional quantum nearest-neighbor Ising model with uniformly distributed random site-dependent spin-spin coupling given by the Hamiltonian 
\begin{equation}
    H = -\sum_{n=1} J_n \sigma_n^z \sigma_{n+1}^z - h_z\sum_{n=1}^N \sigma_n^z - h_x\sum_{n=1}^N \sigma_n^x.
    \label{H}
\end{equation}
There are three parameters appearing in the Hamiltonian~\eqref{H}: $J_n$ is the random spin-spin coupling assuming its value from a probability density function uniformly distributed between $-0.5$ and $0.5$. $h_z$ and $h_x$ are the longitudinal- and transverse-field strengths, which take on a constant value. $\sigma_n^{x}, \; \sigma_n^{z}$ are the Pauli matrices acting on the lattice site $n\in\{1,\ldots,N\}$, with $N$ the total number of sites in the chain. Periodic boundary conditions are considered in the following.
The system for $t<0$ is prepared in the ground state of the Hamiltonian~\eqref{H} at vanishing longitudinal field $h_z$ and spin-spin coupling $J_n$ for each $n$. This leads to the ground state
\begin{equation}\label{psi_0}
    |\psi_0 \rangle = |\psi(t=0) \rangle = |\rightarrow_1 \ldots \rightarrow_N \rangle.
\end{equation}
The quench in the system is performed at $t=0$ when the transverse field $h_x$ is switched off while the longitudinal one ($h_z$) and the spin-spin coupling $J_n$ are switched on. 

\section{Dynamical quantum phase transitions} \label{critical_exponent}
The particular quench introduced in Sec.~\ref{model} has already been studied in the past~\cite{halimeh2019dynamical} and it was observed that close to the critical time of the emerging DQPT, the rate function could be consistent with a power-law behavior of the kind: $|\lambda(t) - \lambda(t_c)| \sim |t - t_c|^\alpha$. One of the main goals of this work is to numerically provide an accurate estimate of the critical exponent $\alpha$. We achieve this because, for the quench considered, the Loschmidt amplitude can be written as a complex partition function of the classical random Ising chain.\cite{ScalingHeyl,PhysRevB.97.174303}
\subsection{Complex Ising partition function}
To see this, consider the initial state~\eqref{psi_0}, which can be written as an equally weighted superposition of eigenstates of the $z$-basis:
$| \psi_0 \rangle = 2^{-N/2} \sum_{s^z} | s^z \rangle$, where $|s^z \rangle$ is of the form: $|s_1,\ldots,s_N \rangle$, with $s_n = \uparrow,\downarrow$. As a consequence, replacing this expression of the initial state into the definition of the Loschmidt amplitude, and noticing that the final Hamiltonian is diagonal in the $z$-basis, all the interference terms vanish ($ \langle (s^{z})'| H | s^z \rangle = 0$ if $| s^z \rangle \neq | (s^{z})' \rangle$) and thus only the diagonal elements remain. Therefore, the Loschmidt amplitude can be recast into the complex partition function
\begin{align}\label{LA_1_main}
	\mathcal{L}(t) 
    =\frac{1}{2^N} \Tr \; e^{-iHt}.
\end{align}
In the uniform limit of such a problem, transfer matrix techniques allow for an exact solution of the free energy in the thermodynamic limit $N \rightarrow \infty$.\cite{huang2009,sachdev2007}
A few changes have to be taken into account at the level of the transfer matrix when computing the Loschmidt amplitude since we are dealing with a time-evolved state\cite{andraschko2014} and a random site-dependent parameter in the Hamiltonian. The final result yields
\begin{equation} \label{L_main}
     \mathcal{L}(t) = \frac{1}{2^N} \Tr(K_1\ldots K_N),
\end{equation}
where $K_n$ is the transfer matrix describing the interactions between two neighboring sites. 
In order to obtain Eq.~\eqref{L_main}, we define

\begin{align}
K(\sigma_n, \sigma_{n+1}) = e^{iJ_nt \sigma_n \sigma_{n+1} + \frac{it}{2}\left(h_n \sigma_n + h_{n+1} \sigma_{n+1}  \right)},
\end{align}
where, for the sake of notational brevity, we have omitted the $z$ superscript in the associated Pauli matrices. Accordingly, the final expression of $\mathcal{L}(t)$ in Eq.~\eqref{LA_1_main} can be written as
\begin{align}\nonumber
	\mathcal{L}(t) = \frac{1}{2^N} \sum_{\sigma_1=\pm1} \ldots 
	\sum_{\sigma_N=\pm1 }  &K(\sigma_1, \sigma_2) K(\sigma_2, \sigma_3)\times\ldots \\\label{LA_2_main}
	\times &K(\sigma_{N-1}, \sigma_N) K(\sigma_{N}, \sigma_1).
\end{align}
Considering $K(\sigma_n, \sigma_{n+1})$ as entries of the $2 \times 2$ matrix $K_n$ (see Appendix~\ref{largeN}) and introducing two states $\sigma^+$ and $\sigma^-$ defined as
\begin{equation}
\begin{aligned}
\sigma^+ = \left( \begin{array}{c}
1  \\
0  \end{array} \right), \;\;
\sigma^- = \left( \begin{array}{c}
0   \\
1  \end{array} \right)
\end{aligned},
\label{sigma_main}
\end{equation}
it turns out that one can write
\begin{equation}
	\begin{split}
		\mathcal{L}(t) =
  		\frac{1}{2^N} \sum_{\alpha_1=\pm} (\sigma^{\alpha_1})^\intercal K_1  K_2 \ldots  K_N \sigma^{\alpha_1}
		= 
		 \frac{1}{2^N} \Tr \prod_{n=1}^N  K_n.
	\end{split}
\label{LE_2_main}
\end{equation}
\subsection{Critical Exponent}
\begin{figure}[tb!]
	\includegraphics[width=1.\columnwidth]{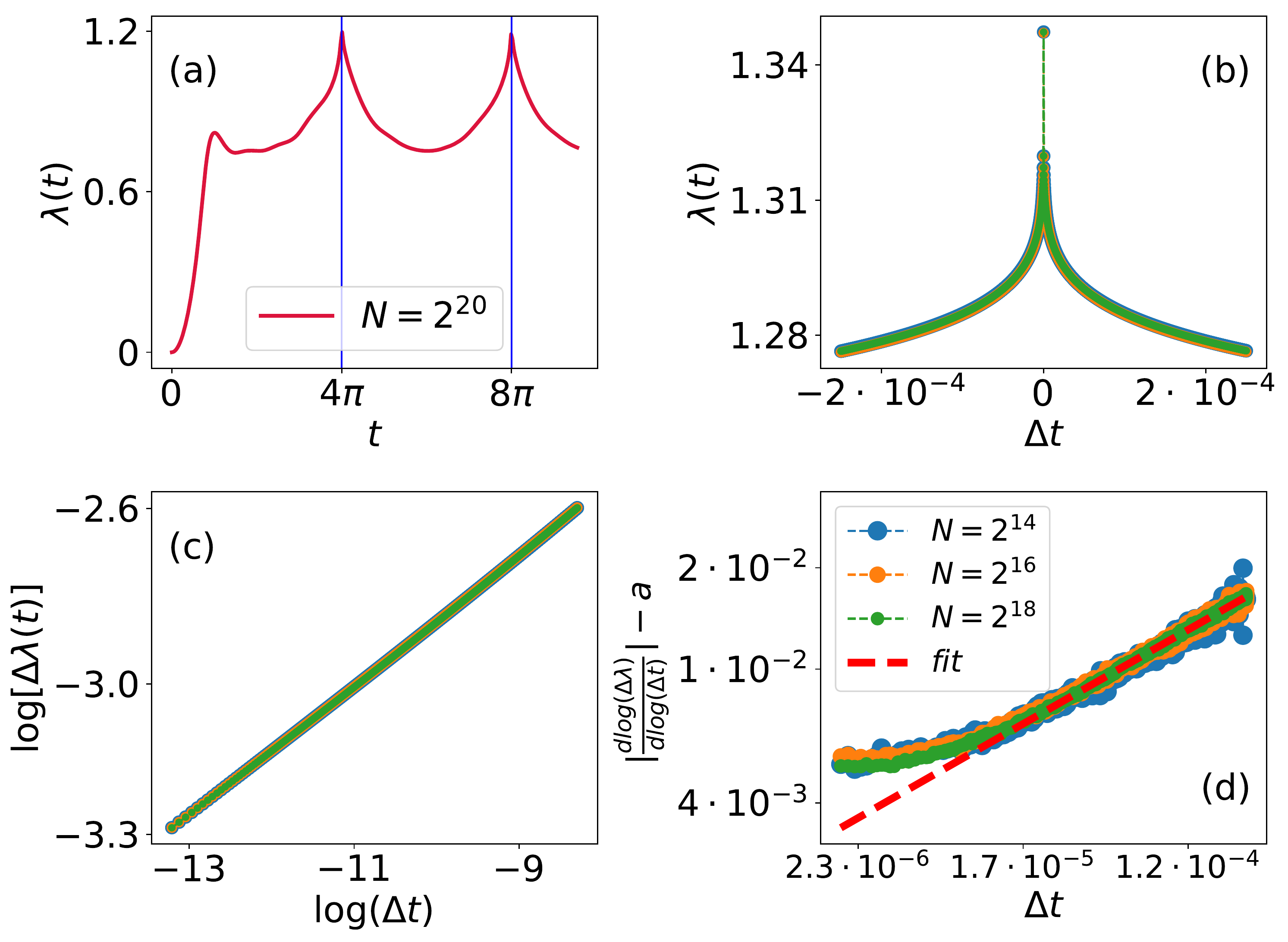}
	\caption{(a) Effective free energy $\lambda$ as a function of time for system size $N=2^{20}$, obtained with $h_z=0.25$. (b) Effective free energy $\lambda$ as a function of $\Delta t = t-t_c$ in the vicinity of the critical time $t_c=4\pi$. The results are shown for three different system sizes: $N=2^{14},2^{16},2^{18}$. At the critical time $t_c$, $\lambda$ exhibits a non-analytical pattern. (c) $\log[\Delta \lambda]$ as a function of $\log[\Delta t]$, where $\Delta \lambda = \lambda(t) - \lambda(t_c)$. The resulting plot looks like a straight line, suggesting a power law relation between $\Delta \lambda$ and $\Delta t$.
	(d) Logarithmic derivative of $\lambda$. It assumes an almost constant value which is the slope of the line in panel (c) and the value of the critical exponent $\alpha$. The red line is the best fit of the form $\mathrm{d} \log|\Delta \lambda| / \mathrm{d} \log|\Delta t|=a+b|\Delta t|^c$ of the points obtained for $N=2^{18}$.}
	\label{fig:rf}
\end{figure}

In our quench protocol, we can reach very large system sizes in performing a spatial decimation RG on the Loschmidt amplitude through merging together two consecutive lattice sites. The result can still be described with a transfer matrix of the same form of the initial problem, but with different parameters.
After the iteration of $N-1$ RG steps, it turns out that the Loschmidt amplitude is given by the product of $N-1$ scalars multiplied by the trace of a $2\times2$ matrix. 
More details are provided in the Appendix~\ref{largeN}.
Exploiting this technique, we can reach very large system sizes($N \sim 2^{20}$), immensely reducing finite-size effects which in general severely undermine the estimation of critical exponents.
The results presented in Fig.~\ref{fig:rf} have been obtained using $h_z=0.25$ and averaging over $3000$ random realizations. We have checked that the same conclusions hold also for other values of the longitudinal field $h_z$.
In Fig.~\ref{fig:rf}(a) we show the effective free energy $\lambda(t)$ as a function of time. In the plot one can see the two times: $4\pi$ and $8\pi$, when $\lambda(t)$ exhibits sharp features---clear consequences of the underlying DQPT of the quench considered.
In Fig.~\ref{fig:rf}(b), we zoom into the vicinity of the critical time observed in panel (a), where on the $x$-axis we consider now the distance to the critical time: $\Delta t= t-t_c$. We show the result for three different system sizes: $N=2^{14},\; 2^{16},\; 2^{18}$. 
In order to understand how the effective free energy scales in the vicinity of the critical time, we focus on Fig.~\ref{fig:rf}(c) where we show $\log|\Delta \lambda|$ vs.~$\log|\Delta t|$, with $\Delta \lambda= \lambda(t) - \lambda(t_c)$.
The almost-straight line observed, suggests a polynomial relation  $|\lambda(t) - \lambda(t_c)| \sim |t-t_c|^\alpha$, where $\alpha$ is the critical exponent. 
Although any system close to the critical point is subjected to severe fluctuations which affect the estimation of the critical exponent, in our case we achieve a good accuracy because of the large system size considered. 
In order to very accurately estimate the value of $\alpha$, we look at the logarithmic derivative of $\lambda(t)$. The result is shown in Fig.~\ref{fig:rf}(d), where we see that $\alpha$ is not yet constant on the whole considered time interval. Nevertheless, upon restricting the range of $\Delta t$ around the critical time, very good convergence is achieved towards a value of 
\begin{align}
\alpha=0.1264(2).
\end{align}
The error here is calculated as the standard deviation of the distance of the points in Fig.~\ref{fig:rf}(d) for $N=2^{18}$ from their best fit with $a+b|\Delta t|^c$. This functional form follows directly from scaling behavior expected at critical points, where $|\Delta\lambda|$ assumes the structure $|\Delta t|^A+g|\Delta t|^B+\ldots$, with real positive coefficients $B>A$, and $g$ a real constant. Thus, for small $\Delta t$, it can be shown that 
\begin{align}\nonumber
\frac{\mathrm{d}\log|\Delta \lambda|}{\mathrm{d}\log|\Delta t|} &=|\Delta t|\frac{\mathrm{d}\log|\Delta \lambda|}{\mathrm{d}|\Delta t|}\\\nonumber
&\approx |\Delta t|\frac{\mathrm{d}\log\big(|\Delta t|^A+g|\Delta t|^B\big)}{\mathrm{d}|\Delta t|}\\\nonumber
&=A+\frac{(B-A)g|\Delta t|^{B-A}}{1+g|\Delta t|^{B-A}}\\
&\sim A+\big(B-A\big)g|\Delta t|^{B-A}.
\end{align}
At very small $|\Delta t|$ (i.e., very close to the critical point), this behavior shows a visible deviation from our data, which we attribute to both finite-size corrections and the inherent imprecision in determining the exact value of the critical point.
The dependency of $\alpha$ on $\Delta t$ is evident when trying to get the critical exponent as the slope of the best linear fitting procedure on the data for $N=2^{18}$ in Fig.~\ref{fig:rf}(c). The result yields $\alpha = 0.13258(7)$, which can be considered as an average of the value of the critical exponent in the whole $\Delta t$-range considered. As expected from Fig.~\ref{fig:rf}(c), this value is affected by the dependency of $\alpha$ on $\Delta t$ and it is slightly higher than the asymptotic result $\alpha=0.1264(2)$ for $\Delta t \rightarrow 0$.
Since the critical exponent is obtained in the limit of $\Delta t \rightarrow 0$, in experiments one should measure the value $\alpha=0.1264$. In fact, this is the numerical result inferred from Fig.~\ref{fig:rf}(d) in the limit $\Delta t \rightarrow 0$. The other result presented in the manuscript: $\alpha = 0.13258(7)$ shows that the numerical value of the critical exponent may depend on the time interval $\Delta t$ considered.

It is worth noting here that Ref.~\onlinecite{halimeh2019dynamical} finds a different value of $\alpha\sim0.2$. The main objective in that work was to show analytically and numerically that disordered models can still exhibit DQPTs, and the maximal number of sites used there was only $N=45000$, much smaller than what we use here ($N=2^{18}$). Moreover, the fit used in Ref.~\onlinecite{halimeh2019dynamical} was a crude one, as is clear from Fig.~3 of that paper. In our present work, extracting the critical exponent was the main objective, and we accordingly directed our numerical efforts towards achieving that. It is this reason why the exponent here is much more accurate than that in Ref.~\onlinecite{halimeh2019dynamical}.

\section{cluster size}\label{cluster_size}
Having shown that quantum quenches in random Ising chains support DQPTs with unconventional exponents, we now aim to explore signatures of these unconventional DQPTs in other physical observables.

\begin{figure}[tb!]
	\includegraphics[width=1.\columnwidth]{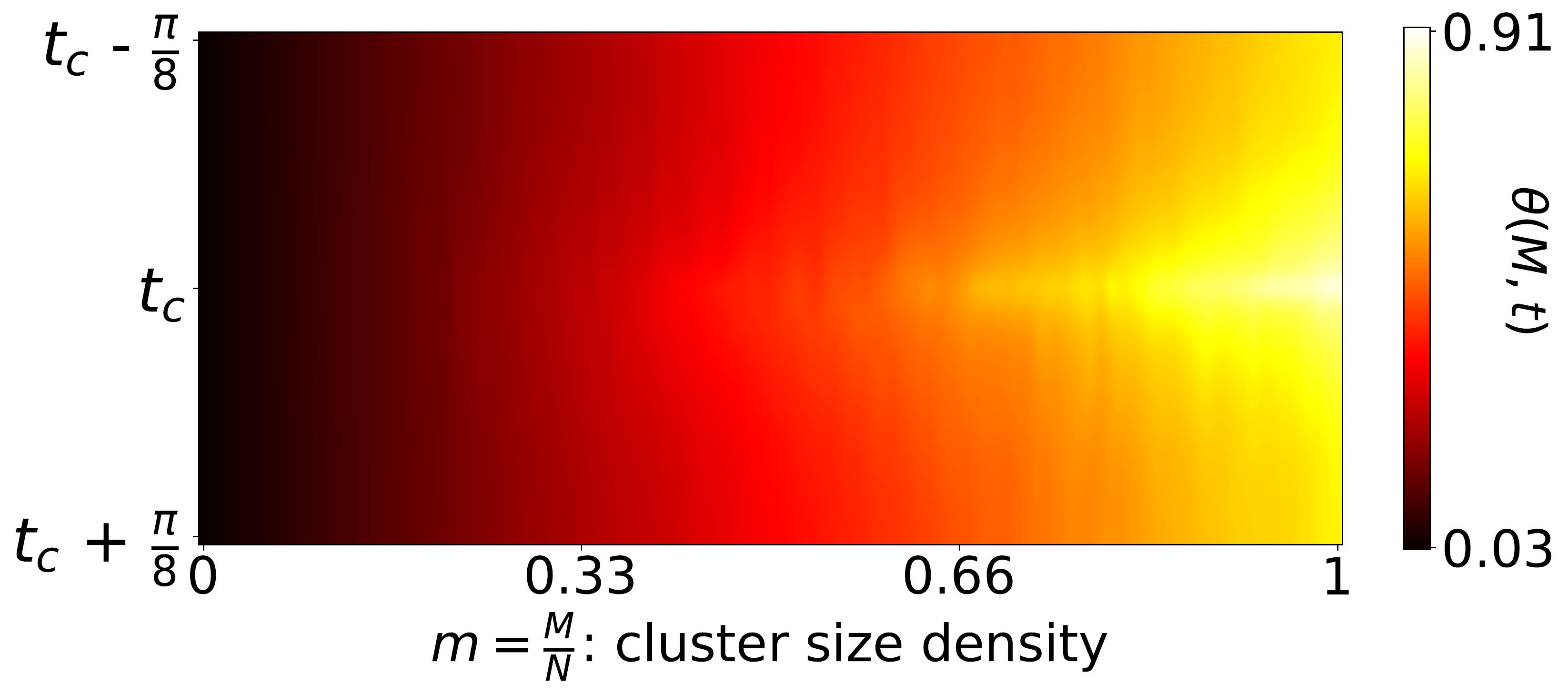}
	\caption{$\theta(M,t) = -\frac{1}{N}\log(|p(M,t)|)$ as a function of time and of cluster size $M$. Around $t_c$, and for large values of $M$, the underlying DQPT affects $\theta(M,t)$, which starts to assume relatively large numbers. This is a consequence of the fact that $\theta(M,t)$ tends to $\lambda$ in the limit of $M\sim N$ and the effective free energy $\lambda$ shows a cusp at $t_c$.}
	\label{fig:p(M)}
\end{figure}

Here, we orient along recent advances in experimental quantum simulation platforms such as in systems of Rydberg atoms,\cite{labuhn2016tunable, barredo2015coherent, kim2018detailed} where the nonequilibrium dynamics of random Ising models as in Eq.~\eqref{H} can be potentially realized,\cite{2017Marcuzzi} as we will discuss in more detail in our concluding discussion in Sec.~\ref{Conclusions}.
Naturally, these architectures provide access to the system properties through projective measurements of spin configurations in a given orientation on the Bloch sphere of each of the qubits.
Here, we will consider the case where the projection is along the $x$-spin direction of each individual atom. 
After multiple measurements such experiments provide naturally access to the statistics of spin configurations, e.g., for our time-evolved state $|\psi(t)\rangle$.
Importantly, this also includes, in principle, the Loschmidt echo itself, as it is nothing but the probability to find at time $t$ the quantum state  $|\psi(t)\rangle$ in the spin configuration $|\psi_0\rangle = |\rightarrow_1 \dots \rightarrow_N\rangle$, see Eq.~\eqref{psi_0}.
Measuring $|\mathcal{L}(t)|^2$, i.e., the probability that the measurement outcome is a single cluster of $\rightarrow$ pointing spins, is only feasible for quantum systems with a limited number of degrees of freedom $N$, as $\mathcal{L}(t)=\exp[-N\lambda(t)]$ is exponentially suppressed.
Smaller clusters of $\rightarrow$ pointing spins can be identified with much less experimental resources.
In the following, we are thus interested in theoretically characterizing the statistics of such clusters and to provide a link to the underlying DQPTs in our setup.
To be specific, a cluster of size $M$ refers to a spatial region with $M$ consecutive spins aligned in the positive $x$-direction, while the two spins at the edges of this string are pointing along the negative $x$-direction, e.g., $\leftarrow_1, \prod_{n=2}^{M+1} \rightarrow_n, \leftarrow_{M+2}$.
For this reason we introduce the on-site projectors
\begin{align}
    \hat{p}_n^{\rightarrow} &= |\rightarrow_n \rangle \langle \rightarrow_n|,\\
    \hat{p}_n^{\leftarrow} &= |\leftarrow_n \rangle \langle \leftarrow_n|,
\end{align}
onto the states $|\rightarrow_n \rangle$ and $| \leftarrow_n \rangle$, respectively.
The probability of finding a cluster of size $M$ at an evolution time $t$ in the chain is 

\begin{align}
    p(M,t)=\langle\psi(t)|\hat{P}_M^x|\psi(t)\rangle,
\end{align}
 where 
\begin{align}\nonumber
		\hat{P}_M^x \coloneqq&\, | \leftarrow_N \rangle \langle \leftarrow_N | \hat{P}_M^{\rightarrow} | \leftarrow_{M+1} \rangle \langle \leftarrow_{M+1} | \\\label{local_proj_2}
		=&\, \hat{p}_N^{\leftarrow} \hat{P}_M^{\rightarrow} \hat{p}_{M+1}^{\leftarrow},
\end{align}
with 
\begin{equation}
		\hat{P}_M^{\rightarrow} \coloneqq	\prod_{n=1}^M \hat{p}_n^{\rightarrow}.
		\label{local_proj}
\end{equation}
We consider now $N=2^7$ and we compute 

\begin{align} 
\theta(M,t)= -\frac{1}{N}\log[p(M,t)],
\end{align}
in the vicinity of the critical time. The result is shown in Fig.~\ref{fig:p(M)}, where we see that the underlying DQPT exhibits a marked influence on the probability distribution function $p(M,t)$. In particular, a clear pattern arises for large values of $M$.

In fact, when $M$ approaches $N$, we notice that $\theta(M,t) \rightarrow \lambda(t)$. This fact can be understood by noticing that for $M \sim N$, the contributions of the two projectors $\hat{p}_N^{\leftarrow}$ and $\hat{p}_{M+1}^{\leftarrow}$ become less and less relevant and, therefore, $\theta(M,t)$ approaches the effective free energy $\lambda(t)$. 
This asymptotic equality holds since the Loschmidt echo, given by $|\mathcal{L}(t)|^2$, can be written in terms of the projector in Eq.~\eqref{local_proj}, with $M=N$:
\begin{align}\nonumber
		|\mathcal{L}(t)|^2 &= |\langle \psi_0 | e^{-iHt} | \psi_0 \rangle|^2=\langle \psi(t)|\psi_0 \rangle \langle \psi_0| \psi(t) \rangle \\
		 &=\langle \psi(t)| \prod_{n=1}^N \hat{p}_n^{\rightarrow} |\psi(t) \rangle=\langle \psi(t)|  \hat{P}_N^{\rightarrow} |\psi(t) \rangle.
\end{align}
Consequently, $\theta(M,t)$ must reproduce the non-analytical pattern for $t=t_c$ which is a mark of the DQPT and well visible in the color plot.
Importantly, the nonanalytic structure at $M=N$ controls $\theta(M,t)$ in a large region for $M<N$.

\section{adding a perturbative random longitudinal field} \label{perturbative}
Up to now we have studied the dynamics of the random Ising chain for homogeneous longitudinal fields.
In the following we will now investigate the influence of a weak inhomogeneity, which is also a particular experimental relevance because imperfections can likely induce such random fields such as for Rydberg atoms, see Sec.~\ref{Conclusions} for a more detailed discussion.
%
\begin{figure}[tb!]
	\includegraphics[width=1.\columnwidth]{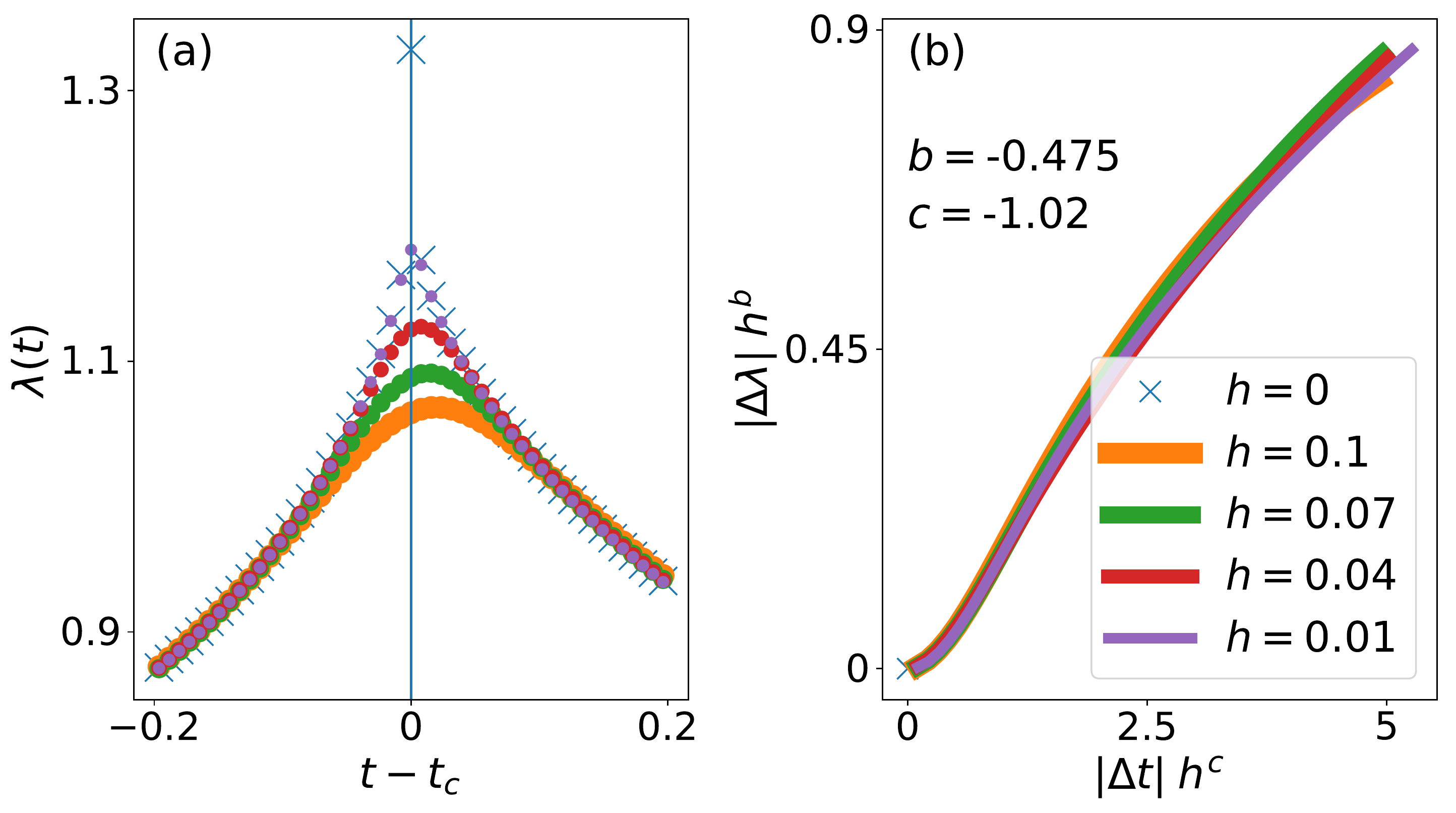}
	\caption{(a) Effective free energy $\lambda(t)$ in the vicinity of the critical time $t_c$ for different values of the longitudinal field $h$: $h=0, \; 0.01, \; 0.04, \; 0.07, \; 0.1$. The system size considered is $N=2^{10}$. In the limit $h=0$ the cusp is clearly visible, while increasing the value of $h$, the pattern becomes more and more smooth.
	(b) $\Delta \lambda$ as a function of $\Delta t$, rescaled by $h^b$ and $h^c$, respectively, for different values of $h$, where $\Delta \lambda = \lambda(t)-\lambda(t_c)$ and $\Delta t = t-t_c$. The curves for different $h$ collapse onto each other when the exponents are $b=-0.475$ and $c=-1.02$.}
	\label{fig:pert_h}
\end{figure}
We therefore now study DQPTs for the extended Hamiltonian
\begin{equation} \label{H_p}
    H_p = H - \sum_{n=1}^N h^R_z(n)\sigma_n^z \, ,
\end{equation}
where the site-dependent random longitudinal field $h^R_z(n)$, with $n=1,\ldots,N$, are independently drawn from a uniform probability distribution centered around 0: $h^R_z(n) \in [-h/2,h/2]$ with $h$ setting the disorder strength. 
We consider the same quench protocol used in Sec.~\ref{critical_exponent}. Using the same methodology introduced in Sec.~\ref{cluster_size}, we determine the effective free energy $\lambda(t)$. We show the results in Fig.~\ref{fig:pert_h}(a) in the vicinity of the critical time for $N=2^{10}$ averaged over around $3000$ random realizations.
While for $h=0$ we see the same DQPT as studied before, this singular feature is smoothed out with increasing $h$.
This is consistent with recent renormalization group considerations, suggesting that the random longitudinal field is a relevant perturbation\cite{PhysRevB.97.174303} meaning that its presence causes the system to be attracted to a different fixed point.
From an alternative perspective, however, the instability of the underlying fixed point implies universality and scaling, even in the presence of this relevant perturbation, which one can make use of.
In this context, the value of $h$ characterizes the distance from the critical point. Consequently we expect to observe some scaling collapse of the curves for different values of $h$ in the vicinity of the critical time upon appropriately rescaling both the distance $\Delta t=t-t_c$ to the critical time as well as the distance $\Delta \lambda=\lambda(t)-\lambda(t_c)$ from the singularity in the effective free energy, by $h^b$ and $h^c$, respectively.
This analysis is shown in Fig.~\ref{fig:pert_h}(b), where the values of the exponents $b$ and $c$ are chosen in such a way to achieve a data collapse for different $h$. It turns out that the exponents are
\begin{equation}
\begin{split}
 b=-0.475,   \\
 c=-1.02.
\end{split}{}    
\end{equation}
Our analysis provides encouraging hopes for a successful experimental realization of such a quench protocol. In fact, although the random longitudinal field smears out the non-analytical pattern of the effective free energy at the critical time, the underlying DQPT at $h=0$ still manifests in the scaling properties of $\lambda(t)$.

\section{Concluding discussion} \label{Conclusions}
In this work we have investigated DQPTs in random Ising models.
Using an exact large-scale renormalization group transformation we have identified with high accuracy the exponent of the DQPT.
As opposed to previously studied cases, where typically integer of mean-field kind of exponents have been found,\cite{Heyl2017Review,wu2019dynamical,wu2020dynamical,wu2020nonequilibrium} we observe in the present model that the exponent is nontrivial.
As already anticipated, the considered nonequilibrium scenario appears feasible within current platforms of Rydberg atoms trapped in optical tweezers. 
In particular, these systems allow to approximately realize the desired target Hamiltonian in Eq.~(\ref{H}).
The  Rydberg interaction generates an effective interaction $V=\sum_{n\not= m} V_{NN} |r_n-r_m|^{-6}(\sigma_n^z+1) (\sigma_m^z+1) $ between the spin degrees of freedom.
Upon tuning the positions $r_n$ of the Rydberg atoms using the optical tweezers, it is possible to realize effective random and inhomogeneous interaction potentials $J_n = V_{NN}/|r_n-r_{n+1}|^6$.\cite{2017Marcuzzi}
In principle, also longer-ranged couplings are present due to the algebraic dependence of the interaction potential.
However, due to the large exponent in the respective power law, further distant couplings are strongly suppressed and can be neglected on the time scales considered in our nonequilibrium setup.
Importantly, the specific form of the Rydberg interactions implies that an inhomogeneous spatial configuration of atoms also leads to inhomogeneous magnetic field contributions $h_n = V_{NN}( |r_{n-1}-r_n|^{-6} + |r_{n}-r_{n+1}|^{-6} )$.
While it might be possible in the future to compensate for these random longitudinal fields with additional locally applied fields, in the short term their presence appears unavoidable and consequently our analysis in Sec.~\ref{perturbative} directly relevant.
Furthermore, Rydberg atoms allow the projective measurements considered for the characterization of cluster sizes in Sec.~\ref{cluster_size}, so that the presented results appear experimentally accessible in current experiments.

For the future, it would be an interesting aspect to use the introduced methodology to study critical exponents in other models such as the one-dimensional Potts model with random couplings, which in the homogeneous case has already been investigated using similar techniques.\cite{2017Karrasch}
A further interesting route might be to study two-dimensional Ising models, where the Loschmidt amplitude can still be identified with a complex classical partition function.\cite{ScalingHeyl}
Such a mapping is still possible for random couplings, where solutions for the classical problem are known~\cite{1985Forgacs,1987Shankar} and might be extended to the nonequilibrium dynamical context.

\begin{acknowledgements}
The authors are grateful to Victor Gurarie for stimulating discussions. This project has received funding from the European Research Council (ERC) under the European Union's Horizon 2020 research and innovation programme (grant agreement No. 853443), and M.~H. further acknowledges support by the Deutsche Forschungsgemeinschaft via the Gottfried Wilhelm Leibniz Prize program. J.C.H. acknowledges support by Provincia Autonoma di Trento, the DFG Collaborative Research Centre SFB 1225 (ISOQUANT), and the ERC Starting Grant StrEnQTh (Project-ID 804305).
	
\end{acknowledgements}

\appendix
\section{Large $N$ for the rate function $\lambda(t)$}\label{largeN}
\subsection{Transfer matrix}
Here, we shall explain in detail how we compute the rate function $\lambda(t)$ for large system sizes.
As mentioned in the main part of the manuscript, the initial state we are considering is the one described in Eq.~\eqref{psi_0}, while the Hamiltonian is given in Eq.~\eqref{H}, with $h_x=0$.
Since this Hamiltonian is diagonal in the $z$-basis,  we are able to write the Loschmidt amplitude in the form of Eq.~\eqref{LA_2_main}. 
The associated terms $K(\sigma_n,\sigma_{n+1})$ form elements of the $2\times2$ matrix 
\begin{align}
	\label{K}
		K_n = C_n
	\begin{pmatrix}
		x_n^{-1} y_n^{-1} & x_n z_n^{-1}   \\
		x_n z_n  & x_n^{-1} y_n
	\end{pmatrix},
\end{align}
with $x_n = e^{-itJ_n}$, $y_n = e^{\frac{-it}{2} (h_n+h_{n+1})}$, and $z_n = e^{\frac{-it}{2} (h_n-h_{n+1})}$. The factor $C_n$ is equal to unity in this case, but its inclusion here is useful, because when performing spatial-RG transformations on the Loschmidt amplitude, its value will change upon application of the RG steps.
We notice that in the uniform limit $J_n = J,\,h_n =h,\,\forall n \in \{1,\ldots,N\}$, the transfer matrix~\eqref{K} yields the well-known result for the partition function of the uniform Ising model.
Using the definition of the vectors $\sigma^{+(-)}$ given in Eq.~\eqref{sigma_main}, we obtain the following identities: 
\begin{align}
	K(1,1)&=(\sigma^+)^\intercal K \sigma^+,\\
	K(1,2)&=(\sigma^+)^\intercal K \sigma^-,\\
	K(2,1)&=(\sigma^-)^\intercal K \sigma^+,\\
	K(2,2)&=(\sigma^-)^\intercal K \sigma^-.
\end{align}
Taking these equalities into account, we can rewrite the Loschmidt amplitude of Eq.~\eqref{LA_2_main} as 
\begin{align}\nonumber
		\mathcal{L}(t)=&\,\frac{1}{2^N}\sum_{\alpha_1=\pm} \ldots \sum_{\alpha_N=\pm } (\sigma^{\alpha_1})^\intercal K_1 \sigma^{\alpha_2} (\sigma^{\alpha_2})^\intercal K_2 \sigma^{\alpha_3} \times \\\nonumber
		&\ldots\times(\sigma^{\alpha_{N-1}})^\intercal K_{N-1}
		\sigma^{\alpha_N} (\sigma^{\alpha_N})^\intercal K_N \sigma^{\alpha_1}\\\nonumber
		=&\,\frac{1}{2^N}\sum_{\alpha_1=\pm} (\sigma^{\alpha_1})^\intercal K_1  \sum_{\alpha_2=\pm }\sigma^{\alpha_2} (\sigma^{\alpha_2})^\intercal  K_2 \sigma^{\alpha_3} \times \\\nonumber
		&\ldots\times(\sigma^{\alpha_{N-1}})^\intercal K_{N-1} \sum_{\alpha_N=\pm } \sigma^{\alpha_N}(\sigma^{\alpha_N})^\intercal K_N \sigma^{\alpha_1}\\\nonumber
  		=&\,\frac{1}{2^N}\sum_{\alpha_1=\pm} (\sigma^{\alpha_1})^\intercal K_1  K_2 \ldots  K_N \sigma^{\alpha_1}\\\label{LE_2}
		=&\,\frac{1}{2^N} \Tr( K_1\ldots K_N) = 
		 \frac{1}{2^N} \Tr\prod_{n=1}^N  K_n,
\end{align}
where we have used the completeness relation of the states $\sigma^{+(-)}$.
\subsection{Spatial RG}
We now perform a spatial RG transformation merging together two consequent lattice sites. For the Ising model, the transfer matrix~\eqref{K} keeps the same form after performing one RG step. Subsequently, we multiply consequent transfer matrices, $K_n K_{n+1}$, and enforce that the result has the same form of the transfer matrix $K$, which is now a function of new couplings $\hat{J}, \hat{h}$. The equality will set the new couplings $\hat{J}, \hat{h}$ (and consequently $\hat{x}$, $\hat{y}$ and $\hat{z}$ ) in terms of the initial parameters $J,h$. This recipe is known as the RG flow equations of the problem.
Explicitly, we have to compute the equation $K_n K_{n+1} = K_{n,n+1}$:
\begin{align} \nonumber
        &C_n C_{n+1}
		\begin{pmatrix}
		x_n^{-1} y_n^{-1} & x_n z_n^{-1}   \\
		x_n z_n  & x_n^{-1} y_n
		\end{pmatrix}
		\begin{pmatrix}
		x_{n+1}^{-1} y_{n+1}^{-1} & x_{n+1} z_{n+1}^{-1}   \\
		x_{n+1} z_{n+1}  & x_{n+1}^{-1} y_{n+1}  
		\end{pmatrix}\\\label{K_RG}
		=&\,\hat{C}_m
		\begin{pmatrix}
		\hat{x}_m^{-1} \hat{y}_m^{-1} & \hat{x}_m \hat{z}_m^{-1}   \\
		\hat{x}_m \hat{z}_m  & \hat{x}_m^{-1} \hat{y}_m  \end{pmatrix}.
\end{align}
The matrix multiplication in Eq.~\eqref{K_RG} leads to the following four equations:
\begin{align}\label{RG_eq_A}
	&\frac{\hat{C}_m}{\hat{y}_m \hat{x}_m}  = \big([x_{n}y_{n}x_{n+1}y_{n+1}]^{-1} + y_{n} z_{n}^{-1} y_{n+1} z_{n+1}\big) C_n C_{n+1} \\\label{RG_eq_B}
	&\hat{C}_m\hat{y}_m \hat{z}_m = \big( y_{n} z_{n} y^{-1}_{n+1} x^{-1}_{n+1} + x_{n} y_{n}^{-1} y_{n+1} z_{n+1} \big) C_n C_{n+1} \\\label{RG_eq_C}
	&\frac{\hat{C}_m\hat{y}_m}{\hat{z}_m}   = \big( y^{-1}_{n} x^{-1}_{n} y_{n+1} z^{-1}_{n+1} + y_{n} z^{-1}_{n} y^{-1}_{n+1} x_{n+1} \big) C_n C_{n+1} \\\label{RG_eq_D}
	&\frac{\hat{C}_m\hat{x}_m}{\hat{y}_m}  = \big( y_{n} z_{n} y_{n+1} z^{-1}_{n+1} + y^{-1}_{n} x_{n} y^{-1}_{n+1} x_{n+1} \big) C_n C_{n+1}.
\end{align}

Dividing Eq.~\eqref{RG_eq_D} by Eq.~\eqref{RG_eq_A}, we get
\begin{align}
	\label{RG_eq_1}
		\hat{x}^2_m = \frac{y_{n} z_{n} y_{n+1} z^{-1}_{n+1} + y^{-1}_{n} x_{n} y^{-1}_{n+1} x_{n+1}}
		{[x_{n}y_{n}x_{n+1}y_{n+1}]^{-1} + y_{n} z_{n}^{-1} y_{n+1} z_{n+1}}.
\end{align}
Similarly, dividing Eq.~\eqref{RG_eq_B} by Eq.~\eqref{RG_eq_C} yields
\begin{align}
\label{RG_eq_2}
\hat{z}^2_m = \frac{ y_{n} z_{n} y^{-1}_{n+1} x^{-1}_{n+1} + x_{n} y_{n}^{-1} y_{n+1} z_{n+1}}
{y^{-1}_{n} x^{-1}_{n} y_{n+1} z^{-1}_{n+1} + y_{n} z^{-1}_{n} y^{-1}_{n+1} x_{n+1}},
\end{align}
and dividing Eq.~\eqref{RG_eq_C} by Eq.~\eqref{RG_eq_A} leads to
\begin{align}\label{RG_eq_3}
\hat{y}^2_m = \frac{y^{-1}_{n} x^{-1}_{n} y_{n+1} z^{-1}_{n+1} + y_{n} z^{-1}_{n} y^{-1}_{n+1} x_{n+1}}
{[x_{n}y_{n}x_{n+1}y_{n+1}]^{-1} + y_{n} z_{n}^{-1} y_{n+1} z_{n+1}} \frac{\hat{z}_m}{\hat{x}_m}.
\end{align}
From Eq.~\eqref{RG_eq_B} we compute
\begin{align}
\label{RG_eq_4}
\hat{C}_m =  \big( y_{n} z_{n} y^{-1}_{n+1} x^{-1}_{n+1} + x_{n} y_{n}^{-1} y_{n+1} z_{n+1} \big)  \frac{C_n C_{n+1}}{\hat{y}_m\hat{z}_m}.
\end{align}

We here make a brief comment on the subscript $m$ of the new couplings appearing in the left-hand side of Eqs.~\eqref{RG_eq_1}--\eqref{RG_eq_4}. At the beginning the chain has $N$ sites. After one RG step, the number of effective sites is halved and consequently also the number of new couplings $\hat{O}$, where the operator $O$ in the previous step can be equal to $C,\;x,\;y,\;z$. We have for example $\hat{O}_1 = O_1 O_2$, $\hat{O}_2 = O_3 O_4$, \ldots$\hat{O}_{N/2} = O_{N-1} O_N$. More generally  $\hat{O}_m = O_n O_{n+1}$ with $2m=n+1$.
The RG analysis presented above turns out to be very useful to compute the Loschmidt amplitude numerically for large system sizes. Indeed, Eq.~\eqref{LE_2} states that the Loschmidt amplitude is given by the product of $N$ matrices. On the other hand, after one RG step the Loschmidt amplitude can be written as a product of $N/2$ matrices: 
\begin{align}
\mathcal{L}(t) =\frac{1}{2^N}  \Tr\prod_{n=1}^{N/2} \hat{K}_n \hat{C}_n.
\end{align}
Performing another RG step we obtain 
\begin{align}
\mathcal{L}(t)=\frac{1}{2^N} \prod_{n=1}^{N/2} \hat{C}_n \Tr\prod_{m=1}^{N/4} \hat{\hat{K}}_m \hat{\hat{C}}_m.
\end{align}
Let us suppose that $N=2^M$, with $M \in \mathbb{N}$. Consequently, after $M-1$ RG steps the effective chain has only one site and the Loschmidt amplitude is given by a product of $N-1$ scalars multiplied by the trace of a single $2 \times 2$ matrix:
\begin{align}\nonumber
		\mathcal{L}(t) = &\,
		\frac{1}{2^N} \prod_{n=1}^{N/2}C^{\text{RG}_1}_n \prod_{m=1}^{N/4} C^{\text{RG}_2}_m \prod_{p=1}^{N/8} C^{\text{RG}_3}_p \ldots\\\nonumber
		&\times\prod_{1=1}^{2} C^{\text{RG}_{M-2}}_q  C^{\text{RG}_{M-1}} \Tr(K^{\text{RG}_{M-1}}) \\\label{LE_scalars}
		=&\, \frac{1}{2^N} \Tr(K^{\text{RG}_{M-1}})  \prod_{n=1}^{N-1}C^{\text{RG}}_n,
	\end{align}
where $C^{\text{RG}_r}_n$ is the coefficient associated to the $n$-site after $r$ RG steps.

\bibliography{BIB_1}

\end{document}